\documentstyle[twocolumn,aps,prb,graphics,floats]{revtex}

\begin{document}

\twocolumn[\hsize\textwidth\columnwidth\hsize\csname
@twocolumnfalse\endcsname

\title{Fractionnally charged excitations in the charge density wave state of a
quarter-filled t-J chain with quantum phonons}

\author{Philippe~Maurel$^{a}$, Marie-Bernadette~Lepetit$^{a}$ 
and Didier~Poilblanc$^{a,b}$}

\address{
$^a$Laboratoire de Physique Quantique \& UMR--CNRS 5626, Universit\'e
Paul Sabatier, F-31062 Toulouse, France
\\
$^b$Theoretische Physik, ETH-H\"onggerberg, CH-8093 Z\"urich
}

\date{\today}

\maketitle

\begin{abstract}
Elementary excitations of the 4k$_F$ charge density
wave state of a quarter-filled strongly correlated electronic
one-dimensional chain are investigated in the presence of
dispersionless quantum optical phonons using Density Matrix
Renormalization Group techniques. Such excitations are shown to 
be topological solitons carrying charge $e/2$ and spin zero. 
Relevance to
the 4k$_F$ charge density wave instability in $\rm (DI-DCNQI)_2A\!g$
or recently discovered in (TMTTF)$_2$X (X=PF$_6$, AsF$_6$) is discussed. 

\end{abstract}

\vskip2pc]



It is well know that one-dimensional (1D) Su-Shrieffer-Heeger
(SSH)~\cite{ssh} or Hubbard-SSH models exhibit exotic elementary
excitations including neutral soliton with spin $1\over 2$, charged
soliton with spin zero ($1\over 2$-filled band)~\cite{chx} as well as
fractionally charged soliton ($1\over 3$- and $1\over 4$-filled
band)~\cite{sol3,sol4}. In these models, the phonons couple to the
electrons via inter-site interactions which lead to an insulating Bond
Order Wave ground state (GS).  In fact, such solitonic excitations are
also generic in commensurate site-centered Charge Density Wave (CDW)
states and, hence, should also exist (in the vicinity of commensurate
fillings) in the case of strong short range electronic repulsion leading to
commensurate $4k_F$ charge instability.
In addition to strong electron-electron correlation, local on-site  
electron-phonon coupling
(to be compared with the inter-site vibration in the SSH model)
is of particular relevance in
systems where the ``site'' represents a complex structure with
internal degrees of freedom. Molecular crystals such as the quasi-1D
charge transfer salts~\cite{review_1D} present this type of
characteristic. 
Interplay between electron-electron and electron-phonon interactions
provides a very rich physics.
For example, several systems have been recently observed to
present transitions towards charge ordered phases where the molecules
of the conducting stack support unequal electron densities, and in
some cases associated relaxation of their interal geometry. This is
for instance the case for the most strongly 1D system of the
Bechgaard-Fabre salts familly, namely $\rm (TMTTF)_2PF_6$ and $\rm
(TMTTF)_2A\!sF_6$; below the Mott localisation temperature $T_\rho$
evidences for an additional transition towards a 4k$_F$
(site-centered) CDW state have been recently
provided by dielectric response measurements~\cite{dielectric}, 
NMR~\cite{nmr_pf6} and anomalous temperature dependence of the X-ray
Bragg peaks~\cite{ravy}. Similar transitions have been
seen in $\rm (BEDT-TTF)_2X$~\cite{bedt} as well as in $\rm
(DI-DCNQI)_2A\!g$~\cite{dcnqi} which exhibits below
$220K$ a $4k_F$ CDW associated with geometry modulations
of the $\rm DI-DCNQI$ molecules.

A pictorial description of a solitonic state can be simply given
assuming e.g.  a quarter-filled strongly correlated chain, in a
$4k_{F}$ CDW state, provided a doubling of the unit cell. In that
case, the GS charge modulation can be parametrized as $\langle n_{i}
\rangle ={1 \over 2} + A_{4k_{F}}cos(4k_{F}r_{i}+\phi)$ where $k_F =
\pi/4$, $A_{4k_{F}}$ is the magnitude and $\phi$ the phase of the
charge oscillation. 
Hence, due to the equivalence between the even and odd sites, the GS
is two fold degenerate ($\phi=0$ and
$\phi=\pi$). A solitonic excitation can be described as a state 
which interpolates
between the two different GS patterns with a slowly monotonically
varying phase $\phi(r_{i})$ from let's say $0$ at $r_{i}\rightarrow
-\infty $ to $\pi$ when $r_{i}\rightarrow +\infty $.
Simple counting arguments show, in fact, that such an excitation
carries a charge $Q=\pm \frac{e}{2}$ and, therefore, can be generated 
by doping the commensurate CDW GS.
  
In this paper, we investigate the role of quantum local phononic
(optical) modes on the formation and on the stability of the solitonic
excitations occuring in an insulating 4k$_F$ CDW phase of a
quarter-filled strongly correlated electronic chain. 
This issue is of
particular interest since a coupling to local phonons 
might affect the physics of the
solitons (such as its width, its interaction, etc...). 
Numerical results will be obtained by the
Density Matrix Renormalization Group (DMRG) method applied to open or
cyclic chain segments carrying no, a single or two solitonic
excitations. 

The following analysis is based on the 1D $t-J-V$-Holstein model at
quarter filling.
This model describing strongly correlated electrons coupled to dispersionless 
phonons can be written as $H=H_{e} + H_{e-ph} $ with
\begin{eqnarray*}
H_{e}= t \sum_{i,\sigma}{c_{i+1,\sigma}^{\dagger}c_{i,\sigma}} + h.c. &+&
J \sum_{i}{{\bf S}_i\cdot {\bf S}_{i+1}} + V\sum_{i}{n_{i}n_{i+1}} \\ 
H_{e-ph}=g\sum_{i}{n_{i}\left(b_{i}^{\dagger}+b_{i}\right)}
&+&\omega\sum_{i}{\left(b_{i}^{\dagger}b_{i}+1/2\right)}
\end{eqnarray*}
where $c_{i,\sigma}^{\dagger }$, $c_{i,\sigma}$ are projected creation
and annihilation operators of electrons of spin $\sigma$ at sites $i$
(doubly occupied sites are projected out, the strong correlation limit
is therefore assumed), $n_i$ is the electron number and ${\bf
S}_i$ is the spin operator at site $i$.
$b_{i}^{\dagger}$ and $b_{i}$ 
are the local phonons creation and annihilation operators. 
The energy scale is fixed by $t=1$. 
Note that the phononic part can be re-written as
$ \omega \left[\left(b_{i}^{\dagger} + n_i{g\over \omega} \right)
\left(b^{\hfill}_{i} + n_i{g\over \omega} \right)\right] $ (apart from
constant terms) showing that
the coupling of the on-site vibrations to the electrons induces 
displacements of the oscillator proportional to the site charge. In
fact, this term mimics the relaxation of the internal geometry 
of a site as a function of its ionicity. 

Before proceeding further, it is interesting to examine the adiabatic
limit $\omega\rightarrow 0$ of our model. 
In that case, absorbing the 
electron-phonon coupling $g$ in the definition of the (classical)
on-site displacement i.e. $g(b_i^\dagger+b_i)\rightarrow \delta_i$,
the phononic part takes the form of a classical
elastic energy $\frac{1}{2}K\sum_i \delta_i^2$.  
The magnitude of the electron-phonon 
coupling is then given by a single parameter 
namely the inverse lattice stiffness $K^{-1}=2g^{2}/\omega$.
Hence, the adiabatic limit is reached assuming the following limits;
$\omega \rightarrow 0$, $g \rightarrow 0$ and $K^{-1}\rightarrow cst$. 
The phase diagram of this model has been investigated 
recently by Riera and Poilblanc~\cite{diagad}. 
It is well know that a quarter-filled infinite-U (i.e. $J=0$) model 
exhibits a $4k_F$ CDW (Mott-Hubbard like) instability only 
when the nearest neighbor (NN) repulsion $V$ exceeds 2~\cite{Haldane}.
This instability is in fact enhanced by the lattice ($\omega=0$)
coupling and the  $4k_F$ CDW phase becomes stable even when
$V<2$ (and $J$ finite) for $K^{-1}$ exceeding a $V$-dependant critical 
value~\cite{diagad}.

The numerical study of the model with quantum phonons using the
infinite system DMRG method~\cite{dmrg} requires an approximate (but reliable)
treatment of the phonon degrees of
freedom~\cite{ph1,ph1p,ph2}. Indeed, an infinite number of phononic
quantum states lives on each site.  In order to render the
calculations feasible, the basis set has been truncated on each site
to the two lowest vibronic states. This choice is physically
reasonable as long as the frequency $\omega$ is not too small since
only the lowest vibronic states are expected to be
involved~\cite{ph2}.  In all cases, we kept $m=216$ states per
renormalized block.  We have chosen parameters like $V=1$ and $J=0.3$
which are generic for strongly correlated 1D materials.  For such
parameters, the adiabatic ground state is a $4k_{F}$ CDW
for $K^{-1} >K^{-1}_{crit}\sim 1.1$.  Ground states and
solitonic states of the system have been investigated as a function of
$\omega$ and $K^{-1}$.

In order to determine the phase diagram at quarter-filling, we 
have computed the charge gap 
$\Delta_{\rho}=E_{0}(2N,N+1)+E_{0}(2N,N-1)-2E_{0}(2N,N)$
(where $E_{0}(N_s,N_e)$ is the ground state energy of $N_e$ electrons on $N_s$
sites). For that purpose, we have used open boundary 
conditions (OBC), systems size up to 50 sites and extrapolated the 
results to the thermodynamic limit.  
Note that OBC are used in gap calculations because the DMRG method performs 
better in this case than with periodic boundary conditions (PBC).
In addition, we have also calculated the charge correlation function
$ c_{n}(j)=\langle(n_{i} - \langle n \rangle)(n_{i+j} 
+ \langle n \rangle )\rangle$ 
where $\langle n \rangle =N_{e}/N_{s}$.


\begin{figure}[htb]
\centerline{\resizebox{8cm}{8cm}{\includegraphics{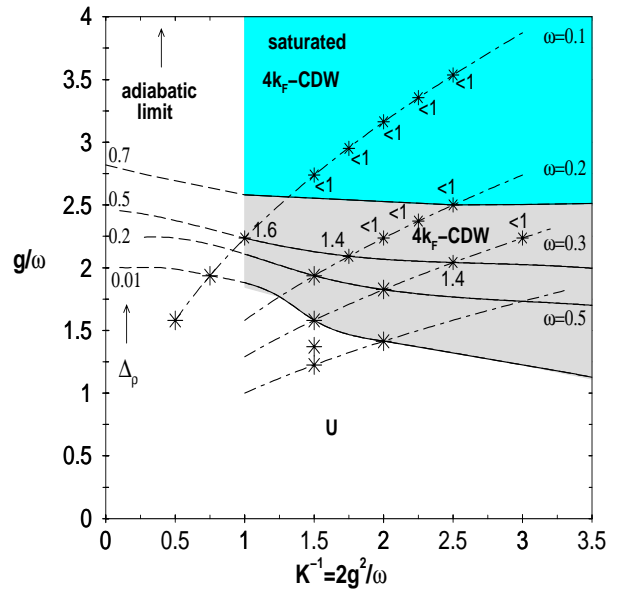}}}
\vspace{5mm}
\caption{
Schematic phase diagram of a ${1 \over 4}$-filled t-J chain
for ${J \over t}=0.3$ and ${V \over t}=1$ as a function of 
$K^{-1}={2g^2 \over \omega}$ and ${g \over \omega}$.
Shaded regions correspond to the insulating CDW phase while 
the uniform phase is labelled by U. 
The darker region corresponds to a nearly fully saturated phase
i.e. $A_{4k_F}\simeq 1/2$. The solid long dashed line 
correspond to iso-gaps curves, the dot-dashed line correspond to 
iso-frequencies curves. 
The numbers correspond to the widths of the solitons $\xi$.
}
\label{fig:diagphase}
\end{figure}

Figure~\ref{fig:diagphase} shows the schematic phase diagram as a
function of $g/\omega$ and $K^{-1}$ exhibiting a $4k_{F}$ CDW
insulating phase and a uniform metallic phase (U).  The insulating
$4k_F$ CDW phase was characterized both by an extrapolated finite
charge gap and long range staggered charge correlations associated to
the finite order parameter $(-1)^{j}c_{n}(j)$.  Special care is needed
for $\omega \rightarrow 0$ since, in this case, the truncation of the
phonon basis is no more adequate and more phonon states are expected
to be excited.  Indeed, $K^{-1}_{crit}\simeq 1.1$ obtained in the
adiabatic approximation~\cite{diagad} does not seem to appear as an
asymptotic limit for the metal-insulator boundary when
$\omega\rightarrow 0$ and $g/\omega \rightarrow +\infty$.  Within our
treatment of the phonons, $K^{-1}_{crit}$ tends towards zero, which
seems inconsistent with the finite value $K^{-1}_{crit}\simeq 1.1$
obtained in the adiabatic approximation~\cite{diagad} for the same $J$
and $V$ values.  Therefore, we
shall restrict in the following analysis to $\omega>0.1$ where we
expect our results to be fully reliable. 

\begin{figure}[htp]
\centerline{\resizebox{6cm}{6cm}{\includegraphics{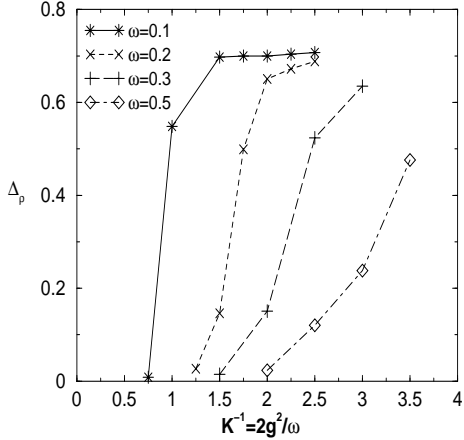}}}
\caption{Charge gap versus $K^{-1}={2g^2\over \omega}$ for different
frequencies (as indicated on the plot).}
\label{fig:gap}
\end{figure}

Let us now discuss the effect of the frequency $\omega$ at fixed
$K^{-1}$: for values of $K^{-1}$ such as the adiabatic ground state is
in the insulating phase, we found that, by increasing $\omega$, the
system becomes metallic.  As one can expect, increasing phonon quantum
fluctuations suppress long-range CDW order.  The opening of the charge
gap characteristic of the metal-insulator transition is fully
consistent with the formation of the CDW as shown in
figure~\ref{fig:gap} and in figure~\ref{fig:diagphase} where the
iso-gap curves are reported.  At intermediate frequencies (let's say
$\omega\geq 0.3)$, we observe a smooth opening of the gap (exponential
like) with a saturation for large value of $K^{-1}$, whereas for
decreasing frequencies the transition seems to become more
abrupt. This behavior is in fact compatible with the first order
character of the metal-insulator transitions seen in the adiabatic
limit~\cite{diagad}.  In contrast, our calculation at finite phonon
frequency would rather be consistent with a Kosterlitz-Thouless type
of gap opening.

By doping (e.g. in electrons) the chain away from the commensurate density of 
$\langle n \rangle=1/2$ one can introduce charged soliton-antisoliton
pairs. Note that solitons naturally appear in pairs since they are
intrinsic topological excitations.
However, in a finite chain, it is possible to enforce the existence of
a single soliton in the GS by assuming an odd number of sites.
For this purpose, we shall deal with odd-length chain 
with $N_{s}=2N+1$ sites and $N_{e}=N+1$ electrons (typically
we choose $N=2p+1$) and PBC. Chains with size up to 43 sites 
have been considered. On the other hand, even-length periodic chains
with $N_{s}=2N$ sites (typically choosing $N=2p+1$) doped with
1 extra electron ($N_{e}=N+1$) of size up to 42 sites 
have been considered to study the behavior of a soliton-antisoliton pair.

\begin{figure}[htp]
\centerline{\resizebox{6cm}{6cm}{\includegraphics{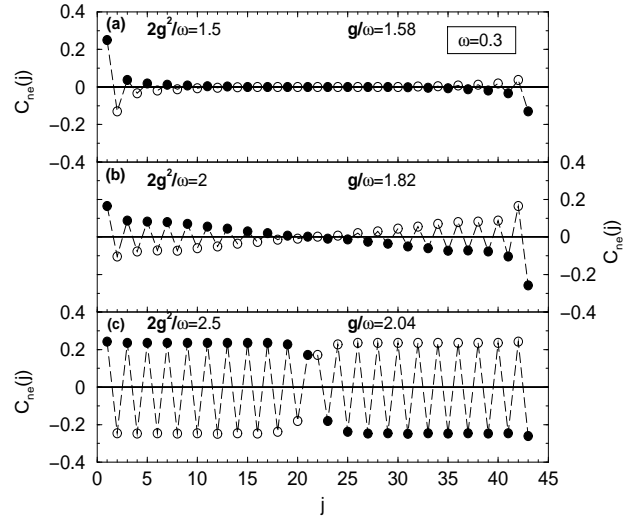}}}
\vspace{1cm}
\caption{Charge correlation function of an odd chain
$N_{s}=43$, $N_{e}=22$ for several couplings $g$.
For convenience data of the same (opposite) 
sign as $(-1)^j$ are shown as full (open) symbols.
For clarity, the soliton has been shifted to the center of the chain.}
\label{fig:corrne1sol}
\end{figure}

The charge-charge correlation in an odd-length chain carrying a single
soliton is shown in Figures~\ref{fig:corrne1sol}(a-c) for different
values of the parameter $g/\omega$ and fixed frequency $\omega=0.3$.
For increasing electron-phonon coupling (or equivalently in this case
for increasing $K^{-1}$), the system evolves from a delocalized state
with no soliton (cf. fig.~\ref{fig:corrne1sol}a) to a state with a
soliton confined on a small number of sites.
(cf. fig.~\ref{fig:corrne1sol}c).  Figure~\ref{fig:corrne1sol}b shows
the intermediate regime where a soliton exists and spreads over a
large number of sites (in fact over all the 43 sites of the largest
chain considered in this work).  In fact, a solitonic excitation
becomes stable and acquires a width and an amplitude at the phase
transition point where the CDW order parameter starts to grow. This
width decreases and the amplitude increases as $K^{-1}$ increases up
to saturation (one inter-site distance for the width and $1/2$ for the
amplitude).  In order to estimate the width of the soliton, one can
fit the staggered charge correlation function with a usual solitonic
function $A\tanh({x-x_{0}\over \xi})$ where $A$ is the long-range CDW
amplitude, $\xi$ is the width of soliton (reported on
figure~\ref{fig:diagphase}) and $x_{0}$ is the location of the center
of the soliton.
When the gap is saturated the soliton is
totally confined to, let say, a single site.  On the contrary, in the
uniform phase, strictly speaking $A\simeq 0$ and the charge
correlation function decays as a power law. It has been argued that,
although phononic quantum fluctuations are present, such state still
belongs to the Luttinger Liquid universality class~\cite{ph1,ph2}. In
that case, the extra $Q=+\frac{e}{2}$ charge is totally spread over
the full chain.  Around the (infinite size) phase transition line
between the uniform and CDW phases, the soliton will appear spread
out over the entire finite system whenever its size (in the infinite
system limit) becomes larger than the actual system size.

In the CDW phase, the DMRG procedure introduces (despite of the PBC)
 a small translation symmetry breaking
(contained in the initial state) which leads to a localization of the
soliton around some arbitrary location $x_0$ along the chain. This
indicates that, in real materials, this type of excitation could very
easily get pinned by impurities or defects. However we still expect
the soliton to be {\it mobile} in a perfectly pure system.

\begin{figure}[htp]
\centerline{\resizebox{6cm}{6cm}{\includegraphics{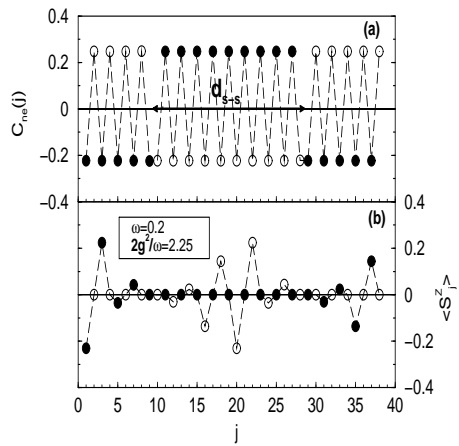}}}
\caption{Charge correlation function (a) and local spin component 
$\big< S_i^Z\big>$ (b) for an even chain with $N_{s}=38$, 
$N_{e}=20$, frequency $\omega =0.2$ and $2g^2/\omega=2.25$.
For convenience, the two sets of data corresponding to the two 
sublattices (``even'' and ``odd'' sites) are shown separately 
as full and open symbols.
}
\label{fig:corr22sol}
\end{figure}

So far, we have imposed the presence of a single soliton in the GS by
a geometrical mean. However, in order to fully prove the stability of such 
excitations one should also consider a situation where at least two of them 
can scatter with each other. For this purpuse, a full extra charge 
$Q=+e$ has been added to a cyclic ring with
an even number of site on top of the CDW GS.
The charge correlation function shown
in figure~\ref{fig:corr22sol}a indicates that a soliton and an
antisoliton well separated from each other appear.
For a fixed system size and different runs,
we have found that the location of the
soliton-antisoliton ``center of mass'' is arbitrary, but the
calculated mean distance
$d_{s-\bar s}$ between the two solitons stays the same. 
For increasing system size $N_{s}$, 
the distance $d_{s-\bar s}$ increases exactly 
like $d_{s-\bar s}={N_{S}\over  2}$. 
This demonstrates that independent solitons and anti-solitons 
are stable in the CDW phase and do not bind into charge $e$ quasiparticles.
Note that the solitonic charge $Q=+\frac{e}{2}$ can be directly
``measured'' by integrating the excess charge over the soliton width.
It is also interesting to notice that previous studies on spin $1\over 2$
solitons in spin-Peierls chains with dynamical phonons~\cite{solsp} 
have demonstrated that soliton-antisoliton bound states cannot exist unless a
two-dimensional coupling is considered. 
Quite generally Ref.~\onlinecite{sol4} predicts solitons to have
either spin zero or spin-1/2.
In fact, the on-site average spin plotted in figure~\ref{fig:corr22sol}b 
shows that, in the present case, each soliton carries no spin.  

We finish this paper by a brief discussion on the possible relevance
to experimental systems. It is clear that with 
on-site CDW associated with differential geometry relaxation of the
$\rm DI-DCNQI$ molecules, the $\rm (DI-DCNQI)_2A\!g$ compound is the
perfect candidate for the present study. Note that under
$5.5K$~\cite{dcnqi2} this compound undergoes a second phase transition
towards an $4k_F$ CDW, $2k_F$ SDW mixed state. As already
mentioned by different authors~\cite{ph2,cdw_coexist}, on-site $4k_F$
CDW have the property to allow simultaneaous $2k_F$ SDW.
Although the metal-insulator instability in 
$\rm (TMTTF)_2X$ (X=PF$_6$, AsF$_6$, etc...) is believed to be
of the Mott-Hubbard type, the recent 
experiments~\cite{dielectric,nmr_pf6,ravy} revealing, at lower
temperatures, a $4k_F$ charge modulation on the $\rm (TMTTF)$ molecules 
suggest that relaxation of the molecules (together with the coupling
to the anions) might play a dominant role.
Therefore, fractionally charged excitations should appear (although a 
dimerization exists along the molecular stacks) and might be revealed 
in e.g. optical
experiments. Note that it has also been theoretically suggested that the
low temperature spin-Peierls phase 
would exhibit, in addition to the lattice tetramerization, a
site-centered $2k_F$ CDW state~\cite{cdw_coexist}.  Recent 
calculations~\cite{integer} suggest that, in the case of
a coexisting $2k_F$ CDW order, two charge $\frac{e}{2}$ solitons would
bind.

\end{document}